\documentclass{article}
\usepackage{preprint,times}
\usepackage[hyphens]{url}
\usepackage{graphicx}
\usepackage{caption}
\usepackage[table]{xcolor}
\usepackage{hyperref}
\usepackage{algorithm}
\usepackage{algorithmic}
\usepackage{booktabs}
\usepackage{colortbl}
\usepackage{amsmath}
\usepackage{amssymb}
\usepackage{tikz}
\usetikzlibrary{positioning,decorations.pathreplacing}

\newcommand{\method}{MeshRepair}
\definecolor{heatmapblue}{HTML}{4C78A8}
\newcommand{\heatmetric}[3]{%
  \pgfmathtruncatemacro{\heatshade}{round(4+16*max(0,min(1,(#1-#2)/(#3-#2))))}%
  \edef\heatfill{\noexpand\cellcolor{heatmapblue!\heatshade}}\heatfill%
  \textcolor{black}{#1}}
\newcommand{\heatdelta}[3]{%
  \pgfmathtruncatemacro{\heatshade}{round(4+16*max(0,min(1,(#1-#2)/(#3-#2))))}%
  \edef\heatfill{\noexpand\cellcolor{heatmapblue!\heatshade}}\heatfill%
  \textcolor{black}{\ensuremath{#1}}}
\newcommand{\heatexcluded}[1]{\cellcolor{black!7}\textcolor{black}{#1}}
\newcommand{\heatexcludeddelta}[1]{%
  \cellcolor{black!7}\textcolor{black}{\ensuremath{#1}}}

\newcommand{\heatdisplay}[4]{%
  \pgfmathtruncatemacro{\heatshade}{round(4+16*max(0,min(1,(#1-#2)/(#3-#2))))}%
  \edef\heatfill{\noexpand\cellcolor{heatmapblue!\heatshade}}\heatfill%
  \textcolor{black}{#4}}
\newcommand{\heatdisplaylow}[4]{%
  \pgfmathtruncatemacro{\heatshade}{round(4+16*max(0,min(1,(#3-#1)/(#3-#2))))}%
  \edef\heatfill{\noexpand\cellcolor{heatmapblue!\heatshade}}\heatfill%
  \textcolor{black}{#4}}
\newcommand{\heatreference}[1]{\cellcolor{black!7}\textcolor{gray}{#1}}

\title{MeshRepair: Autoregressive Topology Repair for Generated Artisan Meshes}

\author{Xiyu Wang$^{1*}$, \enspace Yufei Wang$^{2*\dagger}$, \enspace Ruocheng Wu$^{2}$, \enspace Zhihao Li$^{1,2}$, \enspace Lanqing Guo$^{3}$, \enspace Bihan Wen$^{1\dagger}$ \\[4pt]
{\normalfont $^1$Nanyang Technological University \quad $^2$SparcAI Inc. \quad $^3$University of Texas at Austin} \\
{\normalfont\small $^*$Equal contribution \quad $^\dagger$Corresponding author}}

\preprintfinalcopy
\renewcommand{\headrulewidth}{0pt}
\begin{document}
\maketitle

\begin{figure}[ht]
\centering
\includegraphics[width=\textwidth]{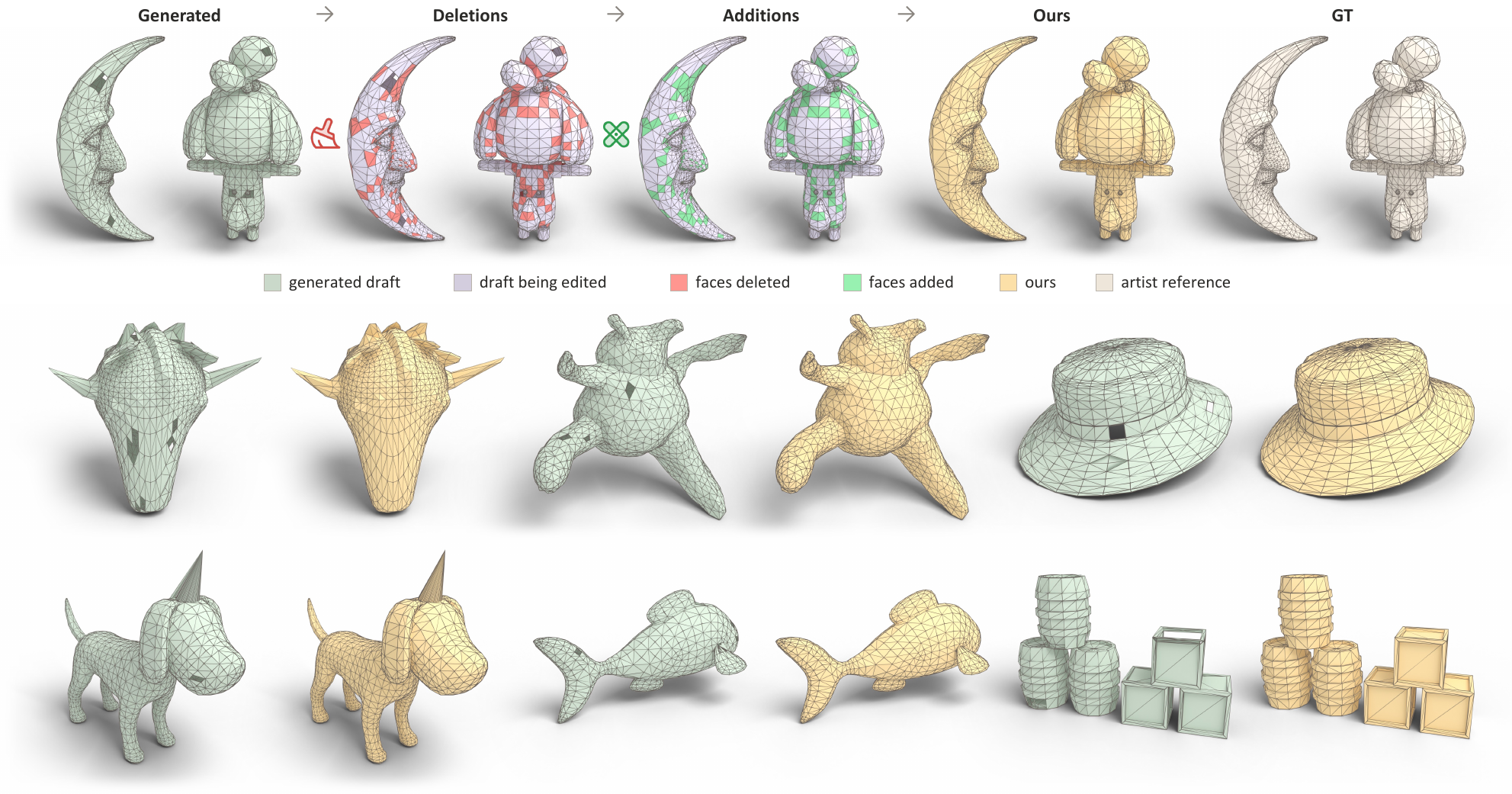}
\caption{\textbf{\method{} repairs generated-mesh topology through face deletion and addition sequences.} Top: \method{} first predicts faces to delete (red) and then faces to add (green), producing the corrected meshes (ours), shown alongside the artist references (GT). Bottom: additional generated drafts and their corrected outputs.}
\label{fig:teaser}
\end{figure}

\begin{abstract}
Recent artisan mesh generation methods produce meshes whose topology comes close to the preferences of human artists. But gaps to artist quality remain because generated meshes commonly contain holes, stray faces, and duplicated faces. Regenerating a mesh from scratch is costly, and a new draw may change the preferred topology entirely or introduce errors elsewhere. In this work we study artisan topology correction. The task is to correct the topology of a generated mesh (the draft) in place, toward the layout an artist would make. We propose \method{} for this task. It is a face-level correction network based on a prefix-LM. It reads the mesh bidirectionally and emits an edit sequence autoregressively. The sequence consists of a deletion phase followed by an addition phase. We propose the novel formulation of the deletion head which picks faces to delete autoregressively and the addition head that selects vertices to form a face to add. The addition head also adds vertices when needed. Untouched faces pass through unchanged, so the corrected mesh keeps the vertices and the layout of the draft. The draft enters as a prefix and only the edits are generated. A correction therefore takes far fewer decoding steps than generating every face again with an autoregressive mesh generator. We benchmark \method{} against repair tools on scripted defects and on held-out drafts from two mesh generators. Across both scripted defects and generated drafts, \method{} improves the quad-equivalent face F1, removes non-manifold faces, and cleans up the topology.
\end{abstract}

\section{Introduction}
\label{sec:intro}

Artisan mesh generation has advanced quickly. Autoregressive models serialize faces into discrete sequences and learn the layouts that artists prefer~\citep{nash2020polygen,siddiqui2024meshgpt,chen2024meshanything,chen2024meshanythingv2,weng2025scaling,tang2024edgerunner,lionar2025treemeshgpt,zhao2025deepmesh,xu2026meshmosaic,wang2026face}. Latent models decode a whole mesh in parallel and reach more complex shapes~\citep{shen2024spacemesh,zhao2026lato,long2026lato2,li2026meshflow,wang2026nexus,wang2026meshcarve}. Structured-latent generators reach high geometric fidelity but output dense surfaces rather than an artist layout~\citep{xiang2025structured,xiang2026native,lai2026lattice,li2025sparc3d}. Their outputs still contain holes, stray faces, duplicated faces, and edges shared by more than two faces. Such defects are local, and the rest of the mesh is what the artist would have modeled. The usual remedy is a new draw from the generator but errors may reappear elsewhere. The topology also changes with every draw, so a layout the user already preferred is lost.

Existing repair tools are rule based. Hole filling in trimesh~\citep{trimesh}, PyMeshLab~\citep{pymeshlab}, and Blender~\citep{blender} closes every boundary loop, although many boundaries in a draft are intended openings. Non-manifold face removal deletes every face at an edge shared by more than two faces, although usually only one of them is wrong. MeshFix~\citep{attene2010meshfix} rebuilds a watertight surface and discards the face layout. None of them can decide which faces the artist intended, because that decision needs the surrounding surface as context and a learned prior of artist layout. MeshPad~\citep{li2025meshpad} is a learned editor, but requires a user-marked region and a target sketch, so it cannot run unattended.

We introduce the task of artisan topology correction and propose \method{}, the first model to perform it. The task is to correct the topology of a generated mesh in place toward the layout an artist would make. It edits the existing draft rather than regenerating the mesh from scratch. The input is the draft. The output is an edit sequence, i.e., the faces to delete and the faces to add. The target is the artist mesh, or reference, while treating the two triangulations of each near-planar quadrilateral as equivalent because the export-time diagonal is arbitrary. We scope the task to topology. \method{} keeps the draft vertices and adds a vertex only where one is missing.

\method{} is a face-level prefix-LM. The draft is canonicalized, i.e., its vertices are quantized to a voxel grid and its faces are sorted, and a small network turns the three vertices of each face into one token~\citep{wang2026face}. A prefix-LM fits the task because the draft is given in full and only the edits are unknown. The face tokens form a prefix that is read bidirectionally, so a missing face can be recognized from the surface around it. Only the edit sequence after the prefix is decoded autoregressively, first all deletions and then all additions, which is far cheaper than regenerating every face. At each deletion step the deletion head ranks the draft faces and removes the top one, until it emits the end-of-del token. At each addition step the addition head names the three vertices of a new face among the draft vertices. Selecting the \textsc{new} token instead predicts the coordinates of a vertex that the draft lacks. We train \method{} with teacher forcing. At inference the finished sequence is applied to the draft, and untouched faces pass through unchanged. By reading the mesh as a prefix, \method{} corrects topology errors automatically, without external inputs such as an editing region, a sketch, a point cloud, or other guidance.

We evaluate \method{} on scripted defects, which a program inserts into artist meshes, and on held-out drafts from two mesh generators. On both, \method{} raises the quad-equivalent face F1, removes nearly all non-manifold edges, and moves vertex valence toward the reference. Repair tools leave the face F1 flat or lower it because difficult cases require identifying which of several faces is wrong---a decision that requires learning a prior over artist-authored mesh topology, rather than indiscriminately closing holes or deleting all candidate faces.

\noindent\textbf{Contributions.}
\begin{itemize}
    \item We introduce artisan topology correction, formulated as predicting topology-editing operations for a draft with defects such as holes, stray faces, and duplicated faces.
    \item We propose \method{}, an efficient prefix-LM with a one-face-one-token prefix and specialized deletion and addition heads, requiring only a fraction of the decoding cost of full-mesh regeneration.
    \item We benchmark \method{} on scripted defects and on generated drafts, and show that it recovers the artist layout in both cases where repair tools fail.
\end{itemize}
\section{Related Work}
\label{sec:related}

\paragraph{Mesh generation.}
Autoregressive mesh models serialize connectivity into discrete sequences: PolyGen uses pointer factorizations~\citep{nash2020polygen}, MeshGPT and MeshAnything generate learned face tokens~\citep{siddiqui2024meshgpt,chen2024meshanything,chen2024meshanythingv2}, BPT, TreeMeshGPT, and MeshMosaic scale to larger meshes~\citep{weng2025scaling,lionar2025treemeshgpt,xu2026meshmosaic}, and EdgeRunner, DeepMesh, and FACE improve scalability and tokenization~\citep{tang2024edgerunner,zhao2025deepmesh,wang2026face}. They represent connectivity exactly but generate a whole mesh sequentially. Latent alternatives such as SpaceMesh, LATO, LATO2, MeshFlow, and MeshCarve encode topology for parallel decoding~\citep{shen2024spacemesh,zhao2026lato,long2026lato2,li2026meshflow,wang2026meshcarve}, and Nexus diffuses vertices and faces directly~\citep{wang2026nexus}. Structured-latent TRELLIS, LATTICE, and Sparc3D achieve high fidelity but produce dense, non-artist surfaces~\citep{xiang2025structured,xiang2026native,lai2026lattice,li2025sparc3d}. Using the draft as a prefix and decoding only edits, we retain efficiency and explicit connectivity without voxel aggregation~\citep{long2026lato2}.

\paragraph{Repair and editing.}
MeshFix enforces watertightness by triangulating every open boundary loop, including openings intentionally left by the artist~\citep{attene2010meshfix}. Besides MeshFix, common 3D modeling libraries also offer standard tools to patch holes and remove stray edges but only target simple defects without any prior of artisan mesh topology. Specifically, trimesh~\citep{trimesh} fills a hole only when its boundary is a triangle or a quadrilateral and it does not patch holes that are larger. PyMeshLab~\citep{pymeshlab} closes holes of any size, but it refuses non-manifold input, so the non-manifold faces must be removed first. Additionally, PyMeshLab can also wrongly detect holes that are in fact proper structures. Non-manifold face removal deletes faces until every edge borders at most two faces but never adds faces, and so opens new holes. Blender~\citep{blender} merges duplicate vertices, removes loose and degenerate geometry, and fills holes. They close every boundary loop, including openings the artist left on purpose, and triangulate it without regard to the artist layout. To this end, several learned methods were proposed. MeshPad~\citep{li2025meshpad} uses separate sketch-conditioned networks for deletion and addition. It is an interactive editor and requires user input. SpaceMesh~\citep{shen2024spacemesh} regenerates the connectivity of a user-marked region, conditioned on a point cloud. In contrast, \method{} requires no user-specified region or sketch and operates fully automatically: a single prefix-LM predicts face deletions and additions from the draft while preserving its unedited connectivity, thereby reducing user effort for the many small, scattered artifacts common in generated meshes.

\section{Method}
\label{sec:method}

\begin{figure*}[t]
\centering
\includegraphics[width=\textwidth]{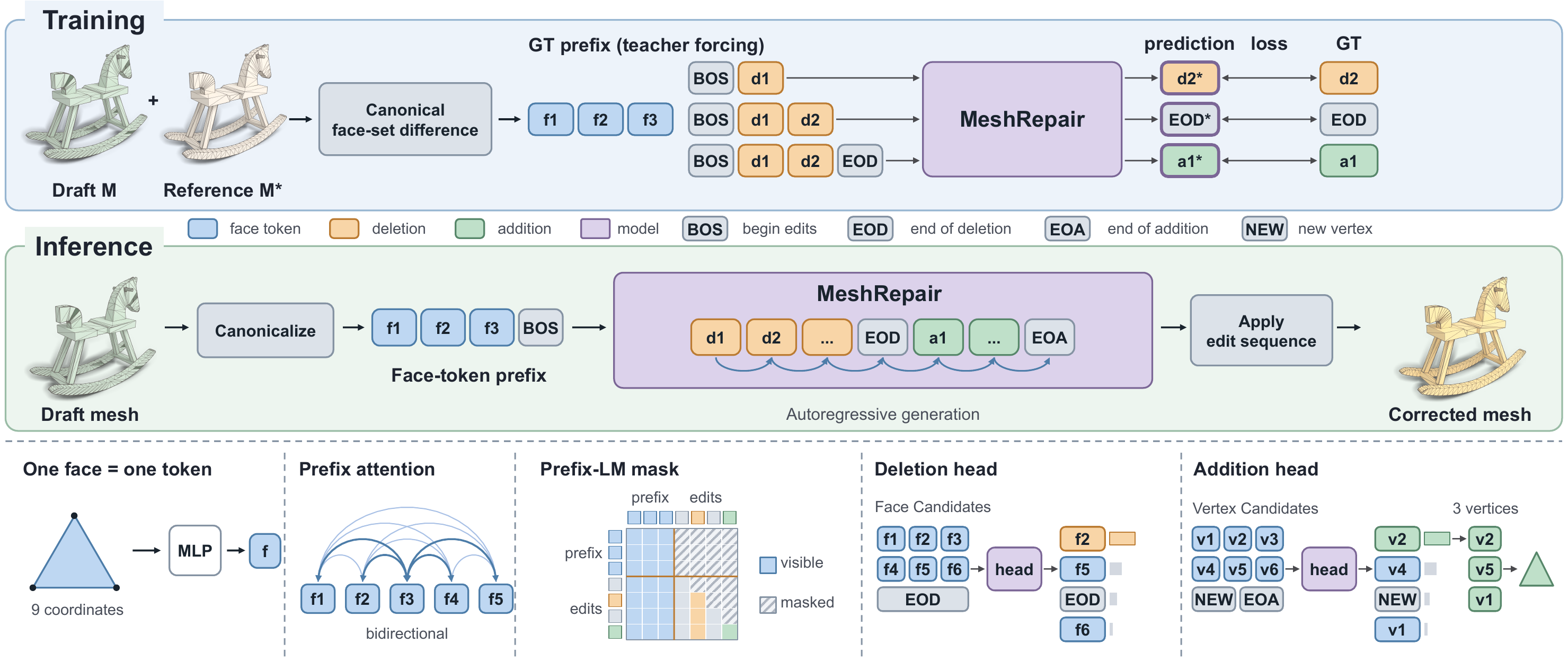}
\caption{\textbf{Overview of \method{}.} Training (top): \method{} receives the draft prefix and the shifted ground-truth edit sequence under a causal mask, so predictions for all edit positions are computed in parallel. The loss compares each prediction (marked $*$) with its ground-truth next token (teacher forcing). Inference (middle): \method{} reads the canonicalized draft as a prefix, generates the deletion phase and then the addition phase autoregressively, and applies the result to the draft. Bottom: face tokenization, bidirectional prefix attention, the prefix-LM mask, and candidate ranking by the two heads (detailed for addition in Algorithm~\ref{alg:add}).}
\label{fig:method}
\end{figure*}
\subsection{Problem Formulation}
\label{sec:method:problem}

We treat topology correction as editing a canonical mesh. Let $M=\{f_1,\dots,f_F\}$ be the canonical draft mesh. Its vertices are quantized to a $512^3$ grid, and each triangular face $f_i$ is represented by the nine integer coordinates of its three vertices, sorted within the face. Duplicate faces are removed, and the remaining faces are sorted. Let $M^\star$ be the canonical reference mesh. \method{} predicts an \emph{edit sequence} $E=(D,A)$, where $D$ indexes draft faces to delete and $A$ holds faces to add. The teacher edit sequence $(D^\star,A^\star)$ is the difference between draft and reference. $D^\star$ indexes the draft faces absent from the reference, and $A^\star$ holds the reference faces absent from the draft. Removing the faces indexed by $D$ from $M$ and adding the faces in $A$ gives the repaired mesh $\mathrm{apply}(M,E)$:
\[
D^\star=\{i:f_i\notin M^\star\},\qquad A^\star=\{f\in M^\star:f\notin M\},\qquad
\mathrm{apply}(M,E)=\{f_i:i\notin D\}\cup A .
\]
Diagonal flips in coplanar quads are excluded from teacher edits (Section~\ref{sec:method:labels}; Figure~\ref{fig:method}).

\subsection{One-Face-One-Token Prefix}
\label{sec:method:tokenize}

Inspired by~\citep{wang2026face}, each face becomes one \emph{face token}: each of its nine coordinates is embedded together with its position among the nine, and a two-layer MLP pools the nine embeddings. The face tokens of the draft form the prefix, and the edit sequence follows it as
\[
\underbrace{f_1\cdots f_F}_{\text{prefix}}\;\langle\textsc{bos}\rangle\;
\underbrace{f_{\delta_1}\cdots f_{\delta_{|D|}}}_{\text{deletion steps}}\;\langle\textsc{eod}\rangle\;
\underbrace{a_1\cdots a_{|A|}}_{\text{addition steps}} ,
\]
where $\delta_1<\dots<\delta_{|D|}$ are the deleted face indices and the added faces follow the canonical order. The begin-of-edits token $\langle\textsc{bos}\rangle$ and the deletion steps predict the next deleted face or the end-of-del token $\langle\textsc{eod}\rangle$. The emitted $\langle\textsc{eod}\rangle$ is fed back, and with the addition steps predicts the next added face or the end-of-add token $\langle\textsc{eoa}\rangle$. Both deletion and addition steps receive the corresponding face-token and face-index embeddings. Segment embeddings mark the prefix and the two phases, and one RoPE counter indexes all positions, so a prefix position is a face index.

\subsection{Prefix-LM Attention}
\label{sec:method:attn}

The prefix is encoded \emph{bidirectionally}, so every prefix token attends to every other prefix token. A hole is a \emph{missing} face and is evident only against the surrounding surface. A causal encoding of the prefix could not attend to the faces that ought to bracket a defect. The edit tokens, i.e., all tokens after the prefix, attend to the entire prefix and causally to earlier edit tokens. Training uses a per-sample block-diagonal prefix-LM mask via FlexAttention, which packs variable-length samples with little padding. Inference uses a KV cache to prefill the prefix bidirectionally and decode edits causally, matching the attention pattern used during training. The backbone is a decoder-only transformer~\citep{vaswani2017attention} (RMSNorm, RoPE, SwiGLU) with GPT-NeoX residual-projection initialization.

\subsection{Deletion Head}
\label{sec:method:del}

The deletion head is a pointer over the prefix faces~\citep{vinyals2015pointer}. For prefix face $i$ we form a key from its final-layer hidden state $h_i$, its face token $e_i$, and a face-index embedding $P(i)$. The query is the hidden state $g_t$ of the current deletion step:
\[
k_i = W_k h_i + W_e e_i + P(i),\quad q_t = W_q g_t,\quad
\ell_{t,i} = \frac{\langle q_t,\,k_i\rangle}{\sqrt{d_p}} ,
\]
where $d_p$ is the key dimension. The softmax runs over the $F$ faces plus a learned key $s$ for $\langle\textsc{eod}\rangle$ with logit $\ell_{t,\varnothing}=\langle q_t,s\rangle/\sqrt{d_p}$. Selecting it ends the deletion phase, and a draft without deletions selects it first. Deletions are emitted in ascending index order, so we apply a \emph{monotonic mask}. At each step only faces indexed above the previously deleted face and $\langle\textsc{eod}\rangle$ are allowed. Training and decoding use the same mask. Feeding the face token of each deleted face rather than only its index lets the addition phase see which faces were removed.

\subsection{Addition Head}
\label{sec:method:add}

For each face addition at step $t$, the addition head constructs a triangular face by selecting its three vertices autoregressively. Let $\mathcal{C}$ denote the current vertex pool, containing draft vertices and vertices generated by earlier additions. For vertex $j$ of the face, the candidates are the vertices in $\mathcal{C}$ and \textsc{new}; selecting \textsc{new} triggers the prediction of a new vertex. A query computed from the hidden state $g_t$ at face-addition step $t$ and the previously selected vertices $c_{<j}=(c_1,\ldots,c_{j-1})$ is matched against each candidate key to produce the selection score $s_j(v)$. Each candidate vertex key combines embeddings of its quantized coordinates with a learned embedding $E_{\mathrm{bnd}}[b(v)]$ of the binary flag $b(v)$ indicating whether the vertex lies on an open mesh boundary; \textsc{new} and \textsc{eoa} use learned keys. We use boundary flags to indicate whether a vertex is on boundaries, providing extra information to the network. For a new vertex, $f_a(i)$ scores coordinate value $i$ conditioned on $g_t$, $c_{<j}$, and the previously predicted coordinates $u_{<a}$. Algorithm~\ref{alg:add} summarizes the addition phase. During both training and decoding, vertices already chosen for the current face are masked out to prevent duplicate vertices. New vertices enter $\mathcal{C}$ and remain selectable for later faces.

\begin{algorithm}[t]
\caption{Addition phase of the addition head.}
\label{alg:add}
\begin{algorithmic}[1]
\STATE $\mathcal{C} \gets$ vertices of the draft
\LOOP
  \STATE $k_v \gets \operatorname{Key}(v,E_{\mathrm{bnd}}[b(v)])$ for $v\in\mathcal{C}$ \COMMENT{coordinate and boundary embeddings}
  \FOR{$j = 1, 2, 3$}
    \STATE $\mathcal{A}_j \gets (\mathcal{C}\setminus\{c_{<j}\})\cup\{\textsc{new}\}$; add $\textsc{eoa}$ if $j=1$ \COMMENT{allowed choices}
    \STATE $c_j \gets \arg\max_{v\in\mathcal{A}_j} s_j(v)$ \COMMENT{$s_j$ selects vertex $j$}
    \STATE \textbf{if} $c_j = \textsc{eoa}$ \textbf{then return}
    \IF{$c_j = \textsc{new}$}
      \FOR{$a = z, y, x$}
        \STATE $u_a \gets \arg\max_{i} f_a(i)$ \COMMENT{$f_a$ scores the $512$ values of coordinate $a$}
      \ENDFOR
      \STATE $c_j \gets (u_z, u_y, u_x)$
      \STATE $\mathcal{C} \gets \mathcal{C}\cup\{c_j\}$ \COMMENT{make the new vertex selectable}
    \ENDIF
  \ENDFOR
  \STATE emit the face $(c_1, c_2, c_3)$ in canonical order
\ENDLOOP
\end{algorithmic}
\end{algorithm}

\subsection{Training and Inference}
\label{sec:method:train}\label{sec:method:labels}

\noindent\textbf{Training.} Artist meshes often contain many quadrilaterals, whose export-time triangulation uses an arbitrary diagonal. We therefore define the target topology up to that choice. When two deleted faces and two added faces split the same four vertices along opposite diagonals, and the four vertices are coplanar within one voxel, the four faces are dropped from the teacher edit sequence. Both triangulations describe the same surface, so \method{} is taught neither diagonal, and the evaluation scores face sets under the same equivalence (Section~\ref{sec:exp:protocol}). Training mixes draft/reference pairs, pairs with part of their edits already applied, and a few identity pairs with an empty edit sequence, which teach \method{} not to edit a correct mesh. Patch hole punching removes further faces in connected patches, which supplies missing faces and new vertices for the addition head. The loss $\mathcal{L}=\mathcal{L}_\mathrm{del}+\mathcal{L}_\mathrm{add}$ sums the teacher-forced cross-entropies of the two heads, each averaged over its own decisions. \method{} has about $2.08$ billion parameters. The base model is trained for $227{,}000$ steps, and the main model fine-tunes it for $14{,}000$ steps on paired generated drafts (Section~\ref{sec:exp:setup}). Appendix~\ref{app:labels} gives implementation details.

\noindent\textbf{Inference.} \method{} prefills the draft, decodes the edit sequence greedily under the same masks as in training, and applies it to the draft once. Winding is restored for export. A few decoding safeguards with no training counterpart, such as caps on the number of edits, are listed in Appendix~\ref{app:labels}.
\begin{figure*}[!t]
\centering
\includegraphics[width=0.85\textwidth]{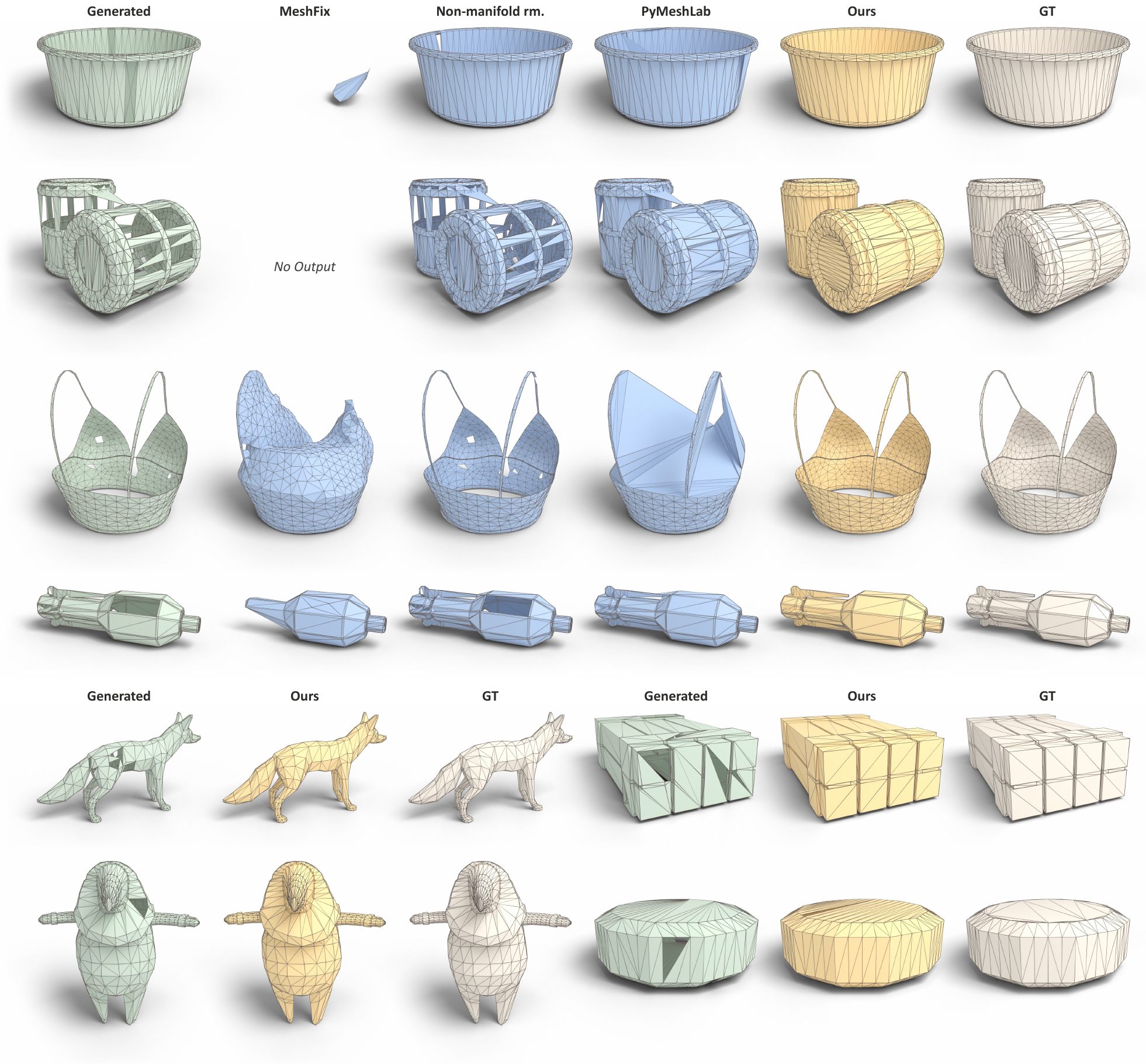}
\caption{\textbf{Selected qualitative corrections.} Top four rows: each generated draft is processed by three repair tools (MeshFix, non-manifold face removal, and PyMeshLab hole closing) and by \method{}, next to the artist reference (GT). Bottom two rows: two drafts per row, each shown as generated draft, \method{}'s correction, and GT. Views and scale are shared within each row. These selected examples illustrate local topology correction. Aggregate performance is reported in Table~\ref{tab:main}.}
\label{fig:qual}
\end{figure*}

\section{Experiments}
\label{sec:exp}

\subsection{Evaluation Protocol}
\label{sec:exp:protocol}
We evaluate topology correction using face-level precision, recall, and F1 rather than relying solely on surface-distance metrics, since the target defects primarily alter face connectivity while leaving vertex positions largely unchanged. Every mesh is canonicalized as in Section~\ref{sec:method:problem} before scoring, so duplicate faces are removed first.

\paragraph{Quad-equivalent face set.}
Artist meshes often contain many quadrilaterals, whose export-time triangulation uses an arbitrary diagonal. Both diagonals describe the same surface and the same modeling intent, so requiring the exact triangles would penalize a correct edit for a choice the artist never made. We therefore identify the two triangulations of a near-planar quadrilateral, a rule we call the quad equivalence. Two faces sharing an edge match the opposite split of the same four vertices when the fourth vertex lies within $\epsilon=1$ voxel of the plane through the other three. No other difference counts as a match, so a missing, extra, or displaced face remains an error. Face precision and recall are the fractions of output and reference faces matched under this rule, and the face F1 combines them. $\Delta$ is the gain in face F1 over the unedited draft. The matching is greedy and bounded as in Appendix~\ref{app:labels}. On a draft where any compared method leaves more than $2{,}048$ unmatched output faces, every method is scored by exact matching instead.

\textbf{Edge F1.} Edge F1 compares undirected edges after snapping each output vertex to a reference vertex in the same voxel or in one of its six face-adjacent voxels. Edges that touch a vertex without a reference counterpart are left out, so the score measures connectivity among shared vertices.

\textbf{Edit accuracy.} Deletion and addition precision and recall ask whether a method made the right edits. The required edits are the exact set differences between draft and reference without the quad equivalence. The edits made are the set differences between draft and output, so the same definition applies to methods that emit no edit sequence. A method that makes no edit of a kind on a pair scores zero precision and recall for it on that pair.

\textbf{Aggregation.} All scores are computed per draft/reference pair and macro-averaged over the stated set of pairs. A reported F1 is therefore the mean of per-pair F1 values rather than the harmonic mean of the averaged precision and recall. Surface distances are reported in Appendix~\ref{app:surface}. Because \method{} keeps the draft vertices, surface distances change little and mainly show whether a method degrades the geometry.

\subsection{Setup}
\label{sec:exp:setup}
\paragraph{Training data.}
Training has two stages. The base model learns from reconstruction pairs, in which a reference mesh is encoded and decoded by a mesh VAE at a continuous range of noise levels. The decoded mesh is the draft, and patch hole punching adds missing faces on top (Section~\ref{sec:method:labels}). The main model then fine-tunes on generated drafts mixed with the reconstruction pairs. MeshCarve~\citep{wang2026meshcarve} and LATO2~\citep{long2026lato2} generate these drafts as connectivity on the vertices of training-split reference meshes. In the generated pairs the draft and reference share one vertex set, so every label is a pure connectivity edit.

\paragraph{Evaluation data.}
We evaluate on two kinds of drafts. The scripted-defect set corrupts held-out artist meshes with two kinds of scripted damage, holes and over-connections: connected patches of faces are removed, and spurious faces are inserted between existing vertices. The generated set uses held-out Objaverse shapes. For each shape, MeshCarve and LATO2 generate connectivity on the reference vertices, and the reference is the quantized artist mesh. Every method is scored on the same drafts. Because the vertices are given, these drafts isolate connectivity errors from vertex placement. Both generators also supply the fine-tuning pairs, so this set tests held-out shapes from generators seen in training.

\paragraph{Baselines.}
We compare against repair tools that need no reference, applied to the same drafts. MeshFix~\citep{attene2010meshfix} rebuilds a manifold surface. Non-manifold face removal deletes faces until no edge is shared by more than two faces, and never adds any. Hole filling in trimesh~\citep{trimesh} and hole closing in PyMeshLab~\citep{pymeshlab} add faces across boundary loops. PyMeshLab first receives the non-manifold removal because it refuses non-manifold input. Blender cleanup~\citep{blender} merges duplicate vertices, removes loose and degenerate geometry, and then fills holes.

\subsection{Main Results}
\label{sec:exp:main}
Table~\ref{tab:main} shows that the repair tools do not recover the artist layout. Hole filling in trimesh, PyMeshLab, and Blender adds faces across boundary loops. Most of the added faces are absent from the reference, so face F1 stays flat or drops on both generators. A boundary in a generated draft is often an intended opening or the side effect of a wrong face elsewhere. A filler that closes every loop cannot tell the two apart. Non-manifold face removal raises precision by deleting faces at edges shared by more than two faces. It cannot tell which of those faces is wrong, so it loses about as much recall as it gains precision. MeshFix rebuilds the surface altogether, which discards the draft connectivity and fails on part of the drafts.

\method{} is the only method that raises face F1 on both generators and on the scripted defects. The gain is led by precision, since \method{} deletes faces the reference does not contain. Its deletion precision is the highest in every block. Unlike non-manifold removal, these deletions are chosen from the whole draft, so face recall is preserved. The gain is larger on LATO2 than on MeshCarve. LATO2 drafts start further from the reference, while MeshCarve drafts are already close and leave less to correct. Edge F1 follows the same ordering, with a clear margin on LATO2 and a small one on MeshCarve. Addition remains the harder half of the task, since added faces are fewer and less precise than deletions. This matches the analysis in Section~\ref{sec:exp:analysis}. On the scripted defects the same pattern appears with a larger gain. \method{} deletes nearly all inserted faces and recovers most of the removed ones, while the repair tools leave face F1 flat or lower it. The scripted defects emulate two error classes that mesh repair surveys document on polygonal models: holes left by surface reconstruction, and redundant faces that produce non-manifold edges~\citep{ju2009fixing,attene2013repairing}. Appendix~\ref{app:surface} reports surface distances. \method{} leaves MeshCarve drafts unchanged within per-pair sampling noise and moves LATO2 drafts closer to the reference. MeshFix, PyMeshLab, and Blender move the surface away from the reference.

\paragraph{Qualitative results.}
Figure~\ref{fig:qual} shows the same behavior on individual drafts. MeshFix returns a remeshed fragment or no mesh at all. Non-manifold removal opens new boundaries where it deletes faces, and hole closing covers openings with fan triangulations. \method{} removes stray and duplicated faces and fills missing ones with faces that follow the artist layout, while the rest of the draft passes through unchanged.

\begin{table*}[t]
\centering
\scriptsize
\setlength{\tabcolsep}{3.4pt}
\caption{\textbf{Topology and editing results.} Macro averages under the protocol in Section~\ref{sec:exp:protocol}. $^\dagger$MeshFix rebuilds the surface, so it deletes nearly every draft face and attains high deletion recall by over-deleting. MeshFix rows average only pairs for which it returns a mesh. Every reported value is shaded on a light blue sequential scale, normalized independently within each data block and column. MeshFix entries are shown in gray and excluded from heatmap normalization because it reconstructs rather than locally edits the input mesh.}
\label{tab:main}
\resizebox{\textwidth}{!}{%
\begin{tabular}{lrrrrrrrrrrr}
\toprule
& \multicolumn{4}{c}{Quad-equivalent face set} & Edge & \multicolumn{3}{c}{Deletion} & \multicolumn{3}{c}{Addition} \\
\cmidrule(lr){2-5}\cmidrule(lr){6-6}\cmidrule(lr){7-9}\cmidrule(l){10-12}
Method & P$\uparrow$ & R$\uparrow$ & F1$\uparrow$ & $\Delta$ & F1$\uparrow$ & P$\uparrow$ & R$\uparrow$ & F1$\uparrow$ & P$\uparrow$ & R$\uparrow$ & F1$\uparrow$ \\
\midrule
\multicolumn{12}{l}{\emph{Scripted defects}} \\
Draft (unedited) & \heatmetric{.901}{.791}{.990} & \heatmetric{.901}{.854}{.973} & \heatmetric{.901}{.837}{.981} & -- & \heatmetric{.922}{.902}{.989} & \heatmetric{0}{0}{.985} & \heatmetric{0}{0}{.988} & \heatmetric{0}{0}{.984} & \heatmetric{0}{0}{.828} & \heatmetric{0}{0}{.702} & \heatmetric{0}{0}{.724} \\
\quad + MeshFix$^\dagger$ & \heatexcluded{.601} & \heatexcluded{.071} & \heatexcluded{.118} & \heatexcludeddelta{-.783} & \heatexcluded{.141} & \heatexcluded{.107} & \heatexcluded{.999} & \heatexcluded{.193} & \heatexcluded{.101} & \heatexcluded{.053} & \heatexcluded{.059} \\
\quad + trimesh hole fill & \heatmetric{.890}{.791}{.990} & \heatmetric{.916}{.854}{.973} & \heatmetric{.902}{.837}{.981} & \heatdelta{+.002}{-.064}{+.081} & \heatmetric{.921}{.902}{.989} & \heatmetric{0}{0}{.985} & \heatmetric{0}{0}{.988} & \heatmetric{0}{0}{.984} & \heatmetric{.502}{0}{.828} & \heatmetric{.133}{0}{.702} & \heatmetric{.203}{0}{.724} \\
\quad + Non-manifold face removal & \heatmetric{.929}{.791}{.990} & \heatmetric{.854}{.854}{.973} & \heatmetric{.890}{.837}{.981} & \heatdelta{-.011}{-.064}{+.081} & \heatmetric{.931}{.902}{.989} & \heatmetric{.427}{0}{.985} & \heatmetric{.340}{0}{.988} & \heatmetric{.375}{0}{.984} & \heatmetric{0}{0}{.828} & \heatmetric{0}{0}{.702} & \heatmetric{0}{0}{.724} \\
\quad + PyMeshLab close holes & \heatmetric{.791}{.791}{.990} & \heatmetric{.889}{.854}{.973} & \heatmetric{.837}{.837}{.981} & \heatdelta{-.064}{-.064}{+.081} & \heatmetric{.902}{.902}{.989} & \heatmetric{.442}{0}{.985} & \heatmetric{.336}{0}{.988} & \heatmetric{.379}{0}{.984} & \heatmetric{.147}{0}{.828} & \heatmetric{.277}{0}{.702} & \heatmetric{.190}{0}{.724} \\
\quad + Blender cleanup + hole fill & \heatmetric{.867}{.791}{.990} & \heatmetric{.936}{.854}{.973} & \heatmetric{.900}{.837}{.981} & \heatdelta{-.001}{-.064}{+.081} & \heatmetric{.919}{.902}{.989} & \heatmetric{.256}{0}{.985} & \heatmetric{.011}{0}{.988} & \heatmetric{.020}{0}{.984} & \heatmetric{.423}{0}{.828} & \heatmetric{.324}{0}{.702} & \heatmetric{.357}{0}{.724} \\
\quad + \method{} (ours) & \heatmetric{.990}{.791}{.990} & \heatmetric{.973}{.854}{.973} & \heatmetric{.981}{.837}{.981} & \heatdelta{+.081}{-.064}{+.081} & \heatmetric{.989}{.902}{.989} & \heatmetric{.985}{0}{.985} & \heatmetric{.988}{0}{.988} & \heatmetric{.984}{0}{.984} & \heatmetric{.828}{0}{.828} & \heatmetric{.702}{0}{.702} & \heatmetric{.724}{0}{.724} \\
\midrule
\multicolumn{12}{l}{\emph{Generated drafts, MeshCarve}} \\
Draft (unedited) & \heatmetric{.838}{.816}{.873} & \heatmetric{.859}{.842}{.865} & \heatmetric{.848}{.832}{.868} & -- & \heatmetric{.829}{.818}{.831} & \heatmetric{0}{0}{.644} & \heatmetric{0}{0}{.223} & \heatmetric{0}{0}{.305} & \heatmetric{0}{0}{.341} & \heatmetric{0}{0}{.093} & \heatmetric{0}{0}{.130} \\
\quad + MeshFix$^\dagger$ & \heatexcluded{.726} & \heatexcluded{.256} & \heatexcluded{.327} & \heatexcludeddelta{-.521} & \heatexcluded{.333} & \heatexcluded{.467} & \heatexcluded{.727} & \heatexcluded{.538} & \heatexcluded{.098} & \heatexcluded{.009} & \heatexcluded{.016} \\
\quad + trimesh hole fill & \heatmetric{.834}{.816}{.873} & \heatmetric{.863}{.842}{.865} & \heatmetric{.847}{.832}{.868} & \heatdelta{.000}{-.016}{.020} & \heatmetric{.829}{.818}{.831} & \heatmetric{0}{0}{.644} & \heatmetric{0}{0}{.223} & \heatmetric{0}{0}{.305} & \heatmetric{.237}{0}{.341} & \heatmetric{.005}{0}{.093} & \heatmetric{.009}{0}{.130} \\
\quad + Non-manifold face removal & \heatmetric{.860}{.816}{.873} & \heatmetric{.842}{.842}{.865} & \heatmetric{.851}{.832}{.868} & \heatdelta{+.003}{-.016}{.020} & \heatmetric{.829}{.818}{.831} & \heatmetric{.602}{0}{.644} & \heatmetric{.070}{0}{.223} & \heatmetric{.117}{0}{.305} & \heatmetric{0}{0}{.341} & \heatmetric{0}{0}{.093} & \heatmetric{0}{0}{.130} \\
\quad + PyMeshLab close holes & \heatmetric{.816}{.816}{.873} & \heatmetric{.850}{.842}{.865} & \heatmetric{.832}{.832}{.868} & \heatdelta{-.016}{-.016}{.020} & \heatmetric{.818}{.818}{.831} & \heatmetric{.603}{0}{.644} & \heatmetric{.067}{0}{.223} & \heatmetric{.111}{0}{.305} & \heatmetric{.097}{0}{.341} & \heatmetric{.010}{0}{.093} & \heatmetric{.016}{0}{.130} \\
\quad + Blender cleanup + hole fill & \heatmetric{.818}{.816}{.873} & \heatmetric{.863}{.842}{.865} & \heatmetric{.839}{.832}{.868} & \heatdelta{-.009}{-.016}{.020} & \heatmetric{.824}{.818}{.831} & \heatmetric{.319}{0}{.644} & \heatmetric{.008}{0}{.223} & \heatmetric{.015}{0}{.305} & \heatmetric{.161}{0}{.341} & \heatmetric{.008}{0}{.093} & \heatmetric{.014}{0}{.130} \\
\quad + \method{} (ours) & \heatmetric{.873}{.816}{.873} & \heatmetric{.865}{.842}{.865} & \heatmetric{.868}{.832}{.868} & \heatdelta{+.020}{-.016}{.020} & \heatmetric{.831}{.818}{.831} & \heatmetric{.644}{0}{.644} & \heatmetric{.223}{0}{.223} & \heatmetric{.305}{0}{.305} & \heatmetric{.341}{0}{.341} & \heatmetric{.093}{0}{.093} & \heatmetric{.130}{0}{.130} \\
\midrule
\multicolumn{12}{l}{\emph{Generated drafts, LATO2}} \\
Draft (unedited) & \heatmetric{.268}{.264}{.346} & \heatmetric{.334}{.292}{.336} & \heatmetric{.294}{.289}{.340} & -- & \heatmetric{.764}{.754}{.784} & \heatmetric{0}{0}{.916} & \heatmetric{0}{0}{.352} & \heatmetric{0}{0}{.484} & \heatmetric{0}{0}{.149} & \heatmetric{0}{0}{.024} & \heatmetric{0}{0}{.038} \\
\quad + MeshFix$^\dagger$ & \heatexcluded{.245} & \heatexcluded{.077} & \heatexcluded{.101} & \heatexcludeddelta{-.193} & \heatexcluded{.274} & \heatexcluded{.803} & \heatexcluded{.788} & \heatexcluded{.771} & \heatexcluded{.026} & \heatexcluded{.001} & \heatexcluded{.003} \\
\quad + trimesh hole fill & \heatmetric{.268}{.264}{.346} & \heatmetric{.334}{.292}{.336} & \heatmetric{.294}{.289}{.340} & \heatdelta{.000}{-.006}{.046} & \heatmetric{.764}{.754}{.784} & \heatmetric{0}{0}{.916} & \heatmetric{0}{0}{.352} & \heatmetric{0}{0}{.484} & \heatmetric{.025}{0}{.149} & \heatmetric{.000}{0}{.024} & \heatmetric{.001}{0}{.038} \\
\quad + Non-manifold face removal & \heatmetric{.297}{.264}{.346} & \heatmetric{.292}{.292}{.336} & \heatmetric{.294}{.289}{.340} & \heatdelta{.000}{-.006}{.046} & \heatmetric{.764}{.754}{.784} & \heatmetric{.907}{0}{.916} & \heatmetric{.210}{0}{.352} & \heatmetric{.311}{0}{.484} & \heatmetric{0}{0}{.149} & \heatmetric{0}{0}{.024} & \heatmetric{0}{0}{.038} \\
\quad + PyMeshLab close holes & \heatmetric{.280}{.264}{.346} & \heatmetric{.299}{.292}{.336} & \heatmetric{.289}{.289}{.340} & \heatdelta{-.006}{-.006}{.046} & \heatmetric{.754}{.754}{.784} & \heatmetric{.910}{0}{.916} & \heatmetric{.199}{0}{.352} & \heatmetric{.300}{0}{.484} & \heatmetric{.022}{0}{.149} & \heatmetric{.003}{0}{.024} & \heatmetric{.004}{0}{.038} \\
\quad + Blender cleanup + hole fill & \heatmetric{.264}{.264}{.346} & \heatmetric{.334}{.292}{.336} & \heatmetric{.292}{.289}{.340} & \heatdelta{-.002}{-.006}{.046} & \heatmetric{.760}{.754}{.784} & \heatmetric{.460}{0}{.916} & \heatmetric{.009}{0}{.352} & \heatmetric{.017}{0}{.484} & \heatmetric{.028}{0}{.149} & \heatmetric{.001}{0}{.024} & \heatmetric{.002}{0}{.038} \\
\quad + \method{} (ours) & \heatmetric{.346}{.264}{.346} & \heatmetric{.336}{.292}{.336} & \heatmetric{.340}{.289}{.340} & \heatdelta{+.046}{-.006}{.046} & \heatmetric{.784}{.754}{.784} & \heatmetric{.916}{0}{.916} & \heatmetric{.352}{0}{.352} & \heatmetric{.484}{0}{.484} & \heatmetric{.149}{0}{.149} & \heatmetric{.024}{0}{.024} & \heatmetric{.038}{0}{.038} \\
\bottomrule
\end{tabular}%
}
\end{table*}
\subsection{Empirical Analysis}
\label{sec:exp:analysis}

\noindent\textbf{Deletion vs.\ addition.} In every block of Table~\ref{tab:main}, the deletions of \method{} are more precise and more complete than its additions. On LATO2 drafts it deletes at a precision of $.916$, while its additions reach a precision of $.149$ and a recall of $.024$. Deletion picks from a fixed set, the draft faces, and only has to decide when to stop. Addition has a much larger action space. It must pick three compatible vertices in order, decide whether a vertex is missing, and stop without repeating a face. Addition also faces a stricter success criterion: when an added face requires a new vertex, the predicted vertex must match the reference voxel exactly for the face to receive credit. Nevertheless, the qualitative results in Figure~\ref{fig:qual} show that \method{} produces plausible additions that repair local defects. The gain in face F1 therefore comes mostly from deletions, and missing faces remain the harder half of the task.

\noindent\textbf{Correction by defect size.} Table~\ref{tab:severity} bins MeshCarve drafts with $r<1$ by the edit ratio $r$, the number of required edits divided by the reference face count. On MeshCarve the gain in face F1 grows with $r$ over the reported range, from $+.004$ at $r$ between $0.02$ and $0.05$ to $+.035$ at $r$ between $0.5$ and $1$, while deletion precision stays above $.5$ for $r\ge 0.02$. Almost every LATO2 draft has $r\ge 1$, and 22\% of its edges are non-manifold (Appendix~\ref{app:structure}), yet \method{} still raises its face F1 by $.046$. Nearly correct MeshCarve drafts show mild over-correction ($-.005$ at $r<0.02$).

\noindent\textbf{Ablations.}
Table~\ref{tab:ablation}(a) replaces each pointer head with a head that emits face coordinates, at a smaller scale. Coordinates for deletion drop deletion F1 from $.662$ to $.091$, since naming a face by its coordinates is far harder than pointing at it. Coordinates for addition drop addition F1 from $.015$ to $.002$ and leave deletion unchanged. Table~\ref{tab:ablation}(b) repeats the correction on its own output. One round captures nearly all of the benefit. Further rounds change face F1 by at most $.001$ on either generator, so they are not worth their cost.
\begin{table*}[t]
\centering
\footnotesize
\begin{minipage}[t]{0.39\textwidth}
\centering
\setlength{\tabcolsep}{3pt}
\caption{\textbf{Ablations.} (a) Pointer vs.\ coordinate heads at a smaller scale (340M parameters, 30k steps, 200 reconstruction pairs). (b) Face F1 after repeated correction rounds.}
\label{tab:ablation}
\resizebox{\linewidth}{!}{%
\begin{tabular}{lcc}
\toprule
\multicolumn{3}{l}{\emph{(a) Pointer vs.\ coordinate heads}}\\
Deletion / addition head & Del.\ F1 & Add.\ F1 \\
\midrule
pointer / pointer       & \heatmetric{.662}{.091}{.669} & \heatmetric{.015}{.002}{.015} \\
coordinates / pointer   & \heatmetric{.091}{.091}{.669} & \heatmetric{.003}{.002}{.015} \\
pointer / coordinates   & \heatmetric{.669}{.091}{.669} & \heatmetric{.002}{.002}{.015} \\
\midrule
\multicolumn{3}{l}{\emph{(b) Correction rounds}}\\
Rounds & MeshCarve & LATO2 \\
\midrule
0 (draft) & \heatmetric{.848}{.848}{.869} & \heatmetric{.294}{.294}{.341} \\
1 & \heatmetric{.868}{.848}{.869} & \heatmetric{.340}{.294}{.341} \\
2 & \heatmetric{.869}{.848}{.869} & \heatmetric{.341}{.294}{.341} \\
3 & \heatmetric{.868}{.848}{.869} & \heatmetric{.341}{.294}{.341} \\
\bottomrule
\end{tabular}}
\end{minipage}\hfill
\begin{minipage}[t]{0.58\textwidth}
\centering
\setlength{\tabcolsep}{3pt}
\caption{\textbf{Correction by defect size.} Generated drafts binned by the edit ratio $r$, the number of edits the reference requires under the quad equivalence (Section~\ref{sec:exp:protocol}) divided by its face count. LATO2 is pooled over all drafts. Face F1 of the draft, its gain from \method{}, the share of drafts that improve or get worse (\%), and deletion and addition precision and recall.}
\label{tab:severity}
\resizebox{\linewidth}{!}{%
\begin{tabular}{lrrrrcccc}
\toprule
 & & \multicolumn{3}{c}{Face} & \multicolumn{2}{c}{Deletion} & \multicolumn{2}{c}{Addition} \\
\cmidrule(lr){3-5}\cmidrule(lr){6-7}\cmidrule(l){8-9}
Drafts by $r$ & $n$ & F1 & $\Delta$ & $\uparrow$/$\downarrow$ & P & R & P & R \\
\midrule
MeshCarve, $<0.02$ & 70 & .998 & $-$.005 & 24/9 & .180 & .024 & .124 & .002 \\
$0.02$--$0.05$ & 43 & .983 & $+$.004 & 70/16 & .510 & .050 & .357 & .019 \\
$0.05$--$0.15$ & 170 & .950 & $+$.012 & 71/23 & .656 & .128 & .382 & .056 \\
$0.15$--$0.3$ & 279 & .887 & $+$.025 & 79/18 & .714 & .193 & .351 & .075 \\
$0.3$--$0.5$ & 195 & .799 & $+$.026 & 69/29 & .679 & .302 & .360 & .136 \\
$0.5$--$1$ & 147 & .692 & $+$.035 & 69/31 & .696 & .426 & .350 & .187 \\
\midrule
LATO2, all & 938 & .294 & $+$.046 & 83/15 & .916 & .352 & .149 & .024 \\
\bottomrule
\end{tabular}}
\end{minipage}
\end{table*}
\section{Limitations}
\label{sec:limits}

\method{} corrects topology only. It preserves existing vertex coordinates, so coordinate drift, misplaced parts, and coarse shape errors remain the generator's responsibility. It also relies on enough local evidence in the draft; extremely sparse, disconnected, or distorted inputs may exceed its capacity. Addition and new-vertex recovery remain harder than deletion, and the model can delete correct faces or leave holes, raising boundary edges on LATO2 (Appendix~\ref{app:structure}). It therefore guarantees neither watertightness, manifoldness, nor exact reference recovery, although its edit sequence can be restricted to a selected region before conventional cleanup.
\section{Conclusion}
\label{sec:conclusion}
We presented \method{}, a face-level prefix-LM that corrects the topology of generated artisan meshes in place. It reads a draft mesh as a prefix to generate deletion and addition sequences, leaving most faces untouched. It requires no other inputs and costs a fraction of full generation. On scripted defects and on drafts from two mesh generators, \method{} raises the quad-equivalent face F1 where repair tools leave it flat or lower it. It also removes nearly all non-manifold edges. Its precise deletions remove erroneous faces, while plausible face additions restore missing connectivity and improve the corrected topology. We see \method{} as a plug-in stage after any mesh generator that turns an almost right draft into an artist-ready mesh.

\bibliography{references}

\begin{thebibliography}{29}
\providecommand{\natexlab}[1]{#1}
\providecommand{\url}[1]{\texttt{#1}}
\expandafter\ifx\csname urlstyle\endcsname\relax
  \providecommand{\doi}[1]{doi: #1}\else
  \providecommand{\doi}{doi: \begingroup \urlstyle{rm}\Url}\fi

\bibitem[Attene(2010)]{attene2010meshfix}
Marco Attene.
\newblock {A Lightweight Approach to Repairing Digitized Polygon Meshes}.
\newblock \emph{The Visual Computer}, 26\penalty0 (11):\penalty0 1393--1406, 2010.

\bibitem[Attene et~al.(2013)Attene, Campen, and Kobbelt]{attene2013repairing}
Marco Attene, Marcel Campen, and Leif Kobbelt.
\newblock {Polygon Mesh Repairing: An Application Perspective}.
\newblock \emph{ACM Computing Surveys}, 45\penalty0 (2):\penalty0 15:1--15:33, 2013.
\newblock \doi{10.1145/2431211.2431214}.

\bibitem[{Blender Online Community}(2025)]{blender}
{Blender Online Community}.
\newblock {Blender, a {3D} modelling and rendering package}.
\newblock \url{https://www.blender.org}, 2025.

\bibitem[Chen et~al.(2025{\natexlab{a}})Chen, He, Huang, Ye, Chen, Tang, Chen, Cai, Yang, Yu, Lin, and Zhang]{chen2024meshanything}
Yiwen Chen, Tong He, Di~Huang, Weicai Ye, Sijin Chen, Jiaxiang Tang, Xin Chen, Zhongang Cai, Lei Yang, Gang Yu, Guosheng Lin, and Chi Zhang.
\newblock {MeshAnything: Artist-Created Mesh Generation with Autoregressive Transformers}.
\newblock In \emph{International Conference on Learning Representations (ICLR)}, 2025{\natexlab{a}}.

\bibitem[Chen et~al.(2025{\natexlab{b}})Chen, Wang, Luo, Wang, Chen, Zhu, Zhang, and Lin]{chen2024meshanythingv2}
Yiwen Chen, Yikai Wang, Yihao Luo, Zhengyi Wang, Zilong Chen, Jun Zhu, Chi Zhang, and Guosheng Lin.
\newblock {MeshAnything V2: Artist-Created Mesh Generation with Adjacent Mesh Tokenization}.
\newblock In \emph{IEEE/CVF International Conference on Computer Vision (ICCV)}, 2025{\natexlab{b}}.

\bibitem[Dawson-Haggerty et~al.(2019)]{trimesh}
Dawson-Haggerty et~al.
\newblock {trimesh}.
\newblock \url{https://trimesh.org/}, 2019.

\bibitem[Ju(2009)]{ju2009fixing}
Tao Ju.
\newblock {Fixing Geometric Errors on Polygonal Models: A Survey}.
\newblock \emph{Journal of Computer Science and Technology}, 24\penalty0 (1):\penalty0 19--29, 2009.
\newblock \doi{10.1007/s11390-009-9206-7}.

\bibitem[Lai et~al.(2026)Lai, Zhao, Zhao, Liu, Lin, Huang, Guo, and Yue]{lai2026lattice}
Zeqiang Lai, Yunfei Zhao, Zibo Zhao, Haolin Liu, Qingxiang Lin, Jingwei Huang, Chunchao Guo, and Xiangyu Yue.
\newblock {LATTICE: Democratize High-Fidelity 3D Generation at Scale}.
\newblock In \emph{Proceedings of the IEEE/CVF Conference on Computer Vision and Pattern Recognition (CVPR)}, 2026.

\bibitem[Li et~al.(2025{\natexlab{a}})Li, Erko{\c{c}}, Li, Sirigatti, Rosov, Dai, and Nie{\ss}ner]{li2025meshpad}
Haoxuan Li, Ziya Erko{\c{c}}, Lei Li, Daniele Sirigatti, Vladislav Rosov, Angela Dai, and Matthias Nie{\ss}ner.
\newblock {MeshPad: Interactive Sketch-Conditioned Artist-Reminiscent Mesh Generation and Editing}.
\newblock In \emph{IEEE/CVF International Conference on Computer Vision (ICCV)}, 2025{\natexlab{a}}.

\bibitem[Li et~al.(2026)Li, Toisoul, Monnier, Shapovalov, Ranjan, Tan, and Vedaldi]{li2026meshflow}
Weiyu Li, Antoine Toisoul, Tom Monnier, Roman Shapovalov, Rakesh Ranjan, Ping Tan, and Andrea Vedaldi.
\newblock {MeshFlow: Efficient Artistic Mesh Generation via MeshVAE and Flow-based Diffusion Transformer}.
\newblock In \emph{IEEE/CVF Conference on Computer Vision and Pattern Recognition (CVPR)}, 2026.

\bibitem[Li et~al.(2025{\natexlab{b}})Li, Wang, Zheng, Luo, and Wen]{li2025sparc3d}
Zhihao Li, Yufei Wang, Heliang Zheng, Yihao Luo, and Bihan Wen.
\newblock {Sparc3D: Sparse Representation and Construction for High-Resolution 3D Shapes Modeling}.
\newblock In D.~Belgrave, C.~Zhang, H.~Lin, R.~Pascanu, P.~Koniusz, M.~Ghassemi, and N.~Chen (eds.), \emph{Advances in Neural Information Processing Systems}, volume 38, Main Conference, pp.\  118582--118600. Curran Associates, Inc., 2025{\natexlab{b}}.
\newblock \doi{10.52202/085713-3956}.
\newblock URL \url{https://proceedings.neurips.cc/paper_files/paper/2025/file/abe731f5242ea6d582f859164b6bbd3c-Paper-Conference.pdf}.

\bibitem[Lionar et~al.(2025)Lionar, Liang, and Lee]{lionar2025treemeshgpt}
Stefan Lionar, Jiabin Liang, and Gim~Hee Lee.
\newblock {TreeMeshGPT: Artistic Mesh Generation with Autoregressive Tree Sequencing}.
\newblock In \emph{Proceedings of the IEEE/CVF Conference on Computer Vision and Pattern Recognition (CVPR)}, pp.\  26608--26617, 2025.

\bibitem[Long et~al.(2026)Long, Zhao, Lin, Zhang, Guo, Liang, Xu, Hladk{\'y}, Nie{\ss}ner, Hu, and Yang]{long2026lato2}
Hang Long, Tianhao Zhao, Junkai Lin, Youjia Zhang, Huipeng Guo, Rendong Liang, Jiale Xu, Jozef Hladk{\'y}, Matthias Nie{\ss}ner, Yuanming Hu, and Wei Yang.
\newblock {LATO.2: Factorized 3D Mesh Generation with Vertex and Topology Flow}, 2026.

\bibitem[Muntoni \& Cignoni(2021)Muntoni and Cignoni]{pymeshlab}
Alessandro Muntoni and Paolo Cignoni.
\newblock {PyMeshLab}, 2021.

\bibitem[Nash et~al.(2020)Nash, Ganin, Eslami, and Battaglia]{nash2020polygen}
Charlie Nash, Yaroslav Ganin, S.~M.~Ali Eslami, and Peter~W. Battaglia.
\newblock {PolyGen: An Autoregressive Generative Model of 3D Meshes}.
\newblock In \emph{International Conference on Machine Learning (ICML)}, 2020.

\bibitem[Shen et~al.(2024)Shen, Li, Law, Atzmon, Fidler, Lucas, Gao, and Sharp]{shen2024spacemesh}
Tianchang Shen, Zhaoshuo Li, Marc Law, Matan Atzmon, Sanja Fidler, James Lucas, Jun Gao, and Nicholas Sharp.
\newblock {SpaceMesh: A Continuous Representation for Learning Manifold Surface Meshes}.
\newblock In \emph{SIGGRAPH Asia Conference Papers}, 2024.

\bibitem[Siddiqui et~al.(2024)Siddiqui, Alliegro, Artemov, Tommasi, Sirigatti, Rosov, Dai, and Nie{\ss}ner]{siddiqui2024meshgpt}
Yawar Siddiqui, Antonio Alliegro, Alexey Artemov, Tatiana Tommasi, Daniele Sirigatti, Vladislav Rosov, Angela Dai, and Matthias Nie{\ss}ner.
\newblock {MeshGPT: Generating Triangle Meshes with Decoder-Only Transformers}.
\newblock In \emph{IEEE/CVF Conference on Computer Vision and Pattern Recognition (CVPR)}, 2024.

\bibitem[Tang et~al.(2025)Tang, Li, Hao, Liu, Zeng, Liu, and Zhang]{tang2024edgerunner}
Jiaxiang Tang, Zhaoshuo Li, Zekun Hao, Xian Liu, Gang Zeng, Ming-Yu Liu, and Qinsheng Zhang.
\newblock {EdgeRunner: Auto-regressive Auto-encoder for Artistic Mesh Generation}.
\newblock In \emph{International Conference on Learning Representations (ICLR)}, 2025.

\bibitem[Vaswani et~al.(2017)Vaswani, Shazeer, Parmar, Uszkoreit, Jones, Gomez, Kaiser, and Polosukhin]{vaswani2017attention}
Ashish Vaswani, Noam Shazeer, Niki Parmar, Jakob Uszkoreit, Llion Jones, Aidan~N. Gomez, Lukasz Kaiser, and Illia Polosukhin.
\newblock {Attention Is All You Need}.
\newblock In \emph{Advances in Neural Information Processing Systems (NeurIPS)}, 2017.

\bibitem[Vinyals et~al.(2015)Vinyals, Fortunato, and Jaitly]{vinyals2015pointer}
Oriol Vinyals, Meire Fortunato, and Navdeep Jaitly.
\newblock {Pointer Networks}.
\newblock In \emph{Advances in Neural Information Processing Systems (NeurIPS)}, 2015.

\bibitem[Wang et~al.(2026{\natexlab{a}})Wang, Guo, Liu, Zou, Zhang, Quan, Liang, Cao, and Yan]{wang2026face}
Hanxiao Wang, Yuan-Chen Guo, Ying-Tian Liu, Zi-Xin Zou, Biao Zhang, Weize Quan, Ding Liang, Yan-Pei Cao, and Dong-Ming Yan.
\newblock {FACE: A Face-based Autoregressive Representation for High-Fidelity and Efficient Mesh Generation}.
\newblock In \emph{IEEE/CVF Conference on Computer Vision and Pattern Recognition (CVPR)}, 2026{\natexlab{a}}.

\bibitem[Wang et~al.(2026{\natexlab{b}})Wang, Liu, Guo, Feng, Zou, Liang, Zhang, and Cao]{wang2026nexus}
Hanxiao Wang, Ying-Tian Liu, Yuan-Chen Guo, Qi-Yuan Feng, Zi-Xin Zou, Ding Liang, Biao Zhang, and Yan-Pei Cao.
\newblock {Nexus: Native Mesh Generation with Diffusion}.
\newblock \emph{ACM Transactions on Graphics (TOG)}, 45\penalty0 (4), 2026{\natexlab{b}}.
\newblock \doi{10.1145/3811344}.

\bibitem[Wang et~al.(2026{\natexlab{c}})Wang, Wu, Wang, Li, Guo, and Wen]{wang2026meshcarve}
Xiyu Wang, Ruocheng Wu, Yufei Wang, Zhihao Li, Lanqing Guo, and Bihan Wen.
\newblock {MeshCarve: Artisan Mesh Generation with Flow Matching in Compact Latent Spaces}.
\newblock arXiv preprint arXiv:2610.09723, 2026{\natexlab{c}}.

\bibitem[Weng et~al.(2025)Weng, Zhao, Lei, Yang, Liu, Lai, Chen, Liu, Jiang, Guo, Zhang, Gao, and Chen]{weng2025scaling}
Haohan Weng, Zibo Zhao, Biwen Lei, Xianghui Yang, Jian Liu, Zeqiang Lai, Zhuo Chen, Yuhong Liu, Jie Jiang, Chunchao Guo, Tong Zhang, Shenghua Gao, and C.~L.~Philip Chen.
\newblock {Scaling Mesh Generation via Compressive Tokenization}.
\newblock In \emph{Proceedings of the IEEE/CVF Conference on Computer Vision and Pattern Recognition (CVPR)}, pp.\  11093--11103, 2025.

\bibitem[Xiang et~al.(2025)Xiang, Lv, Xu, Deng, Wang, Zhang, Chen, Tong, and Yang]{xiang2025structured}
Jianfeng Xiang, Zelong Lv, Sicheng Xu, Yu~Deng, Ruicheng Wang, Bowen Zhang, Dong Chen, Xin Tong, and Jiaolong Yang.
\newblock {Structured 3D Latents for Scalable and Versatile 3D Generation}.
\newblock In \emph{Proceedings of the IEEE/CVF Conference on Computer Vision and Pattern Recognition (CVPR)}, pp.\  21469--21480, 2025.

\bibitem[Xiang et~al.(2026)Xiang, Chen, Xu, Wang, Lv, Deng, Zhu, Dong, Zhao, Yuan, and Yang]{xiang2026native}
Jianfeng Xiang, Xiaoxue Chen, Sicheng Xu, Ruicheng Wang, Zelong Lv, Yu~Deng, Hongyuan Zhu, Yue Dong, Hao Zhao, Nicholas~Jing Yuan, and Jiaolong Yang.
\newblock {Native and Compact Structured Latents for 3D Generation}.
\newblock In \emph{Proceedings of the IEEE/CVF Conference on Computer Vision and Pattern Recognition (CVPR)}, pp.\  14419--14429, 2026.

\bibitem[Xu et~al.(2026)Xu, Xue, Dong, Wan, Zhu, Li, Dou, Lin, Xin, Liu, Wang, and Komura]{xu2026meshmosaic}
Rui Xu, Tianyang Xue, Qiujie Dong, Le~Wan, Zhe Zhu, Peng Li, Zhiyang Dou, Cheng Lin, Shiqing Xin, Yuan Liu, Wenping Wang, and Taku Komura.
\newblock {MeshMosaic: Scaling Artist Mesh Generation via Local-to-Global Assembly}.
\newblock In \emph{Proceedings of the IEEE/CVF Conference on Computer Vision and Pattern Recognition (CVPR)}, pp.\  20003--20013, 2026.

\bibitem[Zhao et~al.(2025)Zhao, Ye, Wang, Liu, Chen, Wang, and Zhu]{zhao2025deepmesh}
Ruowen Zhao, Junliang Ye, Zhengyi Wang, Guangce Liu, Yiwen Chen, Yikai Wang, and Jun Zhu.
\newblock {DeepMesh: Auto-Regressive Artist-mesh Creation with Reinforcement Learning}.
\newblock In \emph{IEEE/CVF International Conference on Computer Vision (ICCV)}, 2025.

\bibitem[Zhao et~al.(2026)Zhao, Zhang, Long, Zhang, Li, Yang, Zhang, Hladk{\'y}, Nie{\ss}ner, and Yang]{zhao2026lato}
Tianhao Zhao, Youjia Zhang, Hang Long, Jinshen Zhang, Wenbing Li, Yang Yang, Gongbo Zhang, Jozef Hladk{\'y}, Matthias Nie{\ss}ner, and Wei Yang.
\newblock {LATO: 3D Mesh Flow Matching with Structured Topology Preserving Latents}.
\newblock In \emph{International Conference on Machine Learning (ICML)}, 2026.

\end{thebibliography}
\bibliographystyle{preprint}

\clearpage
\appendix
\section{Surface Fidelity}
\label{app:surface}
Table~\ref{tab:surface} reports standard surface distances for every method on the generated drafts. These metrics are largely insensitive to the connectivity errors studied here. A flipped quad diagonal, a duplicated face, or a stray face on a non-manifold edge barely moves the sampled surface. MeshCarve drafts already lie close to the sampling floor, so no method can improve them measurably. \method{} leaves them unchanged within sampling noise. Blender, PyMeshLab, and MeshFix move them away from the reference. LATO2 drafts lie further from the floor, and \method{} improves all four.

\begin{table*}[!ht]
\centering
\small
\setlength{\tabcolsep}{5.0pt}
\caption{\textbf{Surface fidelity on generated drafts.} Mean (median) over each set of pairs, with every mesh normalized by its longest bounding-box edge and 20,000 surface samples per mesh. CD-L1 and CD-L2 average the two directional mean distances and squared distances. HD is the symmetric Hausdorff distance over the samples. NC is the bidirectional mean absolute cosine of nearest-neighbor normals. Every comparable value is shaded on a light blue sequential scale, normalized independently within each generator and column; darker cells indicate better values. The gray reference and MeshFix entries are excluded from heatmap normalization. The reference row scores each reference against a fresh sample of itself, i.e., the expected score from sampling alone. $^\dagger$MeshFix rows average only pairs for which it returns a mesh.}
\label{tab:surface}
\begin{tabular}{lcccc}
\toprule
Method & CD-L1$\times10^{3}\downarrow$ & CD-L2$\times10^{5}\downarrow$ & HD$\times10^{2}\downarrow$ & NC$\uparrow$ \\
\midrule
\multicolumn{5}{l}{\emph{Generated drafts, MeshCarve}} \\
\textcolor{gray}{Reference, resampled (floor)} & \heatreference{5.21 (5.11)} & \heatreference{3.76 (3.33)} & \heatreference{2.00 (1.95)} & \heatreference{.950 (.957)} \\
Draft (unedited) & \heatdisplaylow{5.82}{5.79}{7.10}{5.82 (5.42)} & \heatdisplaylow{8.78}{8.67}{23.74}{8.78 (3.86)} & \heatdisplaylow{4.43}{4.19}{6.99}{4.43 (2.77)} & \heatdisplay{.941}{.928}{.941}{.941 (.950)} \\
\quad + MeshFix$^\dagger$ & \heatexcluded{98.18 (93.99)} & \heatexcluded{2116.55 (1381.47)} & \heatexcluded{31.68 (32.77)} & \heatexcluded{.642 (.604)} \\
\quad + trimesh hole fill & \heatdisplaylow{5.81}{5.79}{7.10}{5.81 (5.40)} & \heatdisplaylow{8.67}{8.67}{23.74}{8.67 (3.85)} & \heatdisplaylow{4.19}{4.19}{6.99}{4.19 (2.62)} & \heatdisplay{.941}{.928}{.941}{.941 (.950)} \\
\quad + Non-manifold face removal & \heatdisplaylow{5.79}{5.79}{7.10}{5.79 (5.41)} & \heatdisplaylow{8.77}{8.67}{23.74}{8.77 (3.92)} & \heatdisplaylow{4.63}{4.19}{6.99}{4.63 (3.01)} & \heatdisplay{.941}{.928}{.941}{.941 (.951)} \\
\quad + PyMeshLab close holes & \heatdisplaylow{7.10}{5.79}{7.10}{7.10 (6.00)} & \heatdisplaylow{23.74}{8.67}{23.74}{23.74 (4.92)} & \heatdisplaylow{6.99}{4.19}{6.99}{6.99 (3.59)} & \heatdisplay{.928}{.928}{.941}{.928 (.939)} \\
\quad + Blender cleanup + hole fill & \heatdisplaylow{6.49}{5.79}{7.10}{6.49 (5.61)} & \heatdisplaylow{16.76}{8.67}{23.74}{16.76 (4.40)} & \heatdisplaylow{5.89}{4.19}{6.99}{5.89 (3.05)} & \heatdisplay{.935}{.928}{.941}{.935 (.943)} \\
\quad + \method{} (ours) & \heatdisplaylow{5.93}{5.79}{7.10}{5.93 (5.36)} & \heatdisplaylow{10.35}{8.67}{23.74}{10.35 (3.88)} & \heatdisplaylow{4.23}{4.19}{6.99}{4.23 (2.42)} & \heatdisplay{.940}{.928}{.941}{.940 (.951)} \\
\midrule
\multicolumn{5}{l}{\emph{Generated drafts, LATO2}} \\
\textcolor{gray}{Reference, resampled (floor)} & \heatreference{5.21 (5.11)} & \heatreference{3.76 (3.33)} & \heatreference{2.00 (1.95)} & \heatreference{.950 (.957)} \\
Draft (unedited) & \heatdisplaylow{7.50}{6.57}{8.38}{7.50 (6.23)} & \heatdisplaylow{21.67}{15.00}{34.07}{21.67 (5.32)} & \heatdisplaylow{6.20}{5.60}{8.77}{6.20 (3.64)} & \heatdisplay{.906}{.901}{.932}{.906 (.924)} \\
\quad + MeshFix$^\dagger$ & \heatexcluded{101.00 (95.48)} & \heatexcluded{2153.03 (1441.43)} & \heatexcluded{33.35 (33.99)} & \heatexcluded{.631 (.602)} \\
\quad + trimesh hole fill & \heatdisplaylow{7.49}{6.57}{8.38}{7.49 (6.23)} & \heatdisplaylow{21.64}{15.00}{34.07}{21.64 (5.32)} & \heatdisplaylow{6.20}{5.60}{8.77}{6.20 (3.65)} & \heatdisplay{.906}{.901}{.932}{.906 (.924)} \\
\quad + Non-manifold face removal & \heatdisplaylow{7.09}{6.57}{8.38}{7.09 (5.94)} & \heatdisplaylow{19.71}{15.00}{34.07}{19.71 (4.83)} & \heatdisplaylow{6.07}{5.60}{8.77}{6.07 (3.51)} & \heatdisplay{.916}{.901}{.932}{.916 (.936)} \\
\quad + PyMeshLab close holes & \heatdisplaylow{8.38}{6.57}{8.38}{8.38 (6.67)} & \heatdisplaylow{34.07}{15.00}{34.07}{34.07 (6.93)} & \heatdisplaylow{8.77}{5.60}{8.77}{8.77 (5.64)} & \heatdisplay{.904}{.901}{.932}{.904 (.921)} \\
\quad + Blender cleanup + hole fill & \heatdisplaylow{8.09}{6.57}{8.38}{8.09 (6.55)} & \heatdisplaylow{28.79}{15.00}{34.07}{28.79 (6.56)} & \heatdisplaylow{7.84}{5.60}{8.77}{7.84 (5.00)} & \heatdisplay{.901}{.901}{.932}{.901 (.916)} \\
\quad + \method{} (ours) & \heatdisplaylow{6.57}{6.57}{8.38}{6.57 (5.70)} & \heatdisplaylow{15.00}{15.00}{34.07}{15.00 (4.64)} & \heatdisplaylow{5.60}{5.60}{8.77}{5.60 (3.19)} & \heatdisplay{.932}{.901}{.932}{.932 (.943)} \\
\bottomrule
\end{tabular}
\end{table*}
\section{Mesh Structure}
\label{app:structure}
Table~\ref{tab:structure} reports structural statistics of the corrected meshes. Artist references are not closed manifolds. About a third of them contain non-manifold edges, and boundary edges make up almost four percent of their edges. Each statistic is therefore read against the reference row rather than as a quantity to maximize or minimize. \method{} removes nearly all non-manifold edges and moves vertex valence toward the reference. On MeshCarve it matches the reference valence on more vertices than any other method. It does not reduce boundary edges. On LATO2 drafts it raises their share, because deletions open boundaries that the additions do not fully close. MeshFix and non-manifold removal reach fully manifold output by rebuilding or deleting connectivity, and hole closing removes boundaries the artist left open.

\begin{table*}[t]
\centering
\small
\setlength{\tabcolsep}{5.0pt}
\caption{\textbf{Mesh structure on generated drafts.} Mean over each set of pairs, computed on canonical face sets. Manifold is the share of meshes without non-manifold edges. Non-manifold edges are shared by more than two faces and boundary edges are used by one face, both as shares of all edges. Valence is the mean vertex valence. Val.\ match is the share of reference vertices whose valence in the output equals their valence in the reference. The gray rows describe the artist references. Every comparable value is shaded on the same light blue sequential scale as Table~\ref{tab:main} by its distance to the reference row, normalized independently within each generator and column. Darker cells are closer to the reference. The gray reference and MeshFix entries are excluded from heatmap normalization. $^\dagger$MeshFix rows average only pairs for which it returns a mesh.}
\label{tab:structure}
\resizebox{\textwidth}{!}{%
\begin{tabular}{lccccc}
\toprule
Method & Manifold (\%) & Non-manif.\ edges (\%) & Bound.\ edges (\%) & Valence & Val.\ match (\%) \\
\midrule
\multicolumn{6}{l}{\emph{Generated drafts, MeshCarve}} \\
\textcolor{gray}{Reference} & \heatreference{67.4} & \heatreference{0.39} & \heatreference{3.69} & \heatreference{5.59} & \heatreference{100.0} \\
Draft (unedited) & \heatdisplaylow{55.5000}{18.7000}{55.7000}{11.9} & \heatdisplaylow{5.0300}{0.2800}{5.0300}{5.42} & \heatdisplaylow{1.2000}{0.0100}{3.4500}{4.89} & \heatdisplaylow{0.0300}{0.0100}{0.1600}{5.62} & \heatdisplaylow{58.8000}{56.5000}{60.9000}{41.2} \\
\quad + MeshFix$^\dagger$ & \heatexcluded{99.8} & \heatexcluded{0.02} & \heatexcluded{0.25} & \heatexcluded{5.68} & \heatexcluded{11.1} \\
\quad + trimesh hole fill & \heatdisplaylow{55.6000}{18.7000}{55.7000}{11.8} & \heatdisplaylow{5.0200}{0.2800}{5.0300}{5.41} & \heatdisplaylow{0.0100}{0.0100}{3.4500}{3.68} & \heatdisplaylow{0.0500}{0.0100}{0.1600}{5.64} & \heatdisplaylow{58.9000}{56.5000}{60.9000}{41.1} \\
\quad + Non-manifold face removal & \heatdisplaylow{32.6000}{18.7000}{55.7000}{100.0} & \heatdisplaylow{0.3900}{0.2800}{5.0300}{0.00} & \heatdisplaylow{2.7500}{0.0100}{3.4500}{6.44} & \heatdisplaylow{0.0200}{0.0100}{0.1600}{5.57} & \heatdisplaylow{58.6000}{56.5000}{60.9000}{41.4} \\
\quad + PyMeshLab close holes & \heatdisplaylow{18.7000}{18.7000}{55.7000}{48.7} & \heatdisplaylow{0.2800}{0.2800}{5.0300}{0.11} & \heatdisplaylow{3.4500}{0.0100}{3.4500}{0.24} & \heatdisplaylow{0.1600}{0.0100}{0.1600}{5.75} & \heatdisplaylow{60.9000}{56.5000}{60.9000}{39.1} \\
\quad + Blender cleanup + hole fill & \heatdisplaylow{55.7000}{18.7000}{55.7000}{11.7} & \heatdisplaylow{4.7800}{0.2800}{5.0300}{5.17} & \heatdisplaylow{1.8500}{0.0100}{3.4500}{1.84} & \heatdisplaylow{0.1100}{0.0100}{0.1600}{5.70} & \heatdisplaylow{60.2000}{56.5000}{60.9000}{39.8} \\
\quad + \method{} (ours) & \heatdisplaylow{22.7000}{18.7000}{55.7000}{44.7} & \heatdisplaylow{0.3000}{0.2800}{5.0300}{0.69} & \heatdisplaylow{1.2700}{0.0100}{3.4500}{4.96} & \heatdisplaylow{0.0100}{0.0100}{0.1600}{5.58} & \heatdisplaylow{56.5000}{56.5000}{60.9000}{43.5} \\
\midrule
\multicolumn{6}{l}{\emph{Generated drafts, LATO2}} \\
\textcolor{gray}{Reference} & \heatreference{67.3} & \heatreference{0.39} & \heatreference{3.67} & \heatreference{5.59} & \heatreference{100.0} \\
Draft (unedited) & \heatdisplaylow{64.7000}{24.3000}{65.1000}{2.6} & \heatdisplaylow{21.5000}{0.1800}{21.5000}{21.89} & \heatdisplaylow{0.6100}{0.6100}{5.9200}{3.06} & \heatdisplaylow{0.3900}{0.0800}{0.4500}{5.98} & \heatdisplaylow{66.6000}{64.2000}{67.4000}{33.4} \\
\quad + MeshFix$^\dagger$ & \heatexcluded{99.7} & \heatexcluded{0.01} & \heatexcluded{2.86} & \heatexcluded{5.61} & \heatexcluded{11.5} \\
\quad + trimesh hole fill & \heatdisplaylow{64.7000}{24.3000}{65.1000}{2.6} & \heatdisplaylow{21.4900}{0.1800}{21.5000}{21.88} & \heatdisplaylow{0.7200}{0.6100}{5.9200}{2.95} & \heatdisplaylow{0.3900}{0.0800}{0.4500}{5.98} & \heatdisplaylow{66.5000}{64.2000}{67.4000}{33.5} \\
\quad + Non-manifold face removal & \heatdisplaylow{32.7000}{24.3000}{65.1000}{100.0} & \heatdisplaylow{0.3900}{0.1800}{21.5000}{0.00} & \heatdisplaylow{5.9200}{0.6100}{5.9200}{9.59} & \heatdisplaylow{0.1200}{0.0800}{0.4500}{5.71} & \heatdisplaylow{64.2000}{64.2000}{67.4000}{35.8} \\
\quad + PyMeshLab close holes & \heatdisplaylow{46.8000}{24.3000}{65.1000}{20.5} & \heatdisplaylow{0.1800}{0.1800}{21.5000}{0.57} & \heatdisplaylow{2.4400}{0.6100}{5.9200}{1.23} & \heatdisplaylow{0.3300}{0.0800}{0.4500}{5.92} & \heatdisplaylow{66.0000}{64.2000}{67.4000}{34.0} \\
\quad + Blender cleanup + hole fill & \heatdisplaylow{65.1000}{24.3000}{65.1000}{2.2} & \heatdisplaylow{21.2000}{0.1800}{21.5000}{21.59} & \heatdisplaylow{2.0100}{0.6100}{5.9200}{1.66} & \heatdisplaylow{0.4500}{0.0800}{0.4500}{6.04} & \heatdisplaylow{67.4000}{64.2000}{67.4000}{32.6} \\
\quad + \method{} (ours) & \heatdisplaylow{24.3000}{24.3000}{65.1000}{43.0} & \heatdisplaylow{0.4300}{0.1800}{21.5000}{0.82} & \heatdisplaylow{3.3700}{0.6100}{5.9200}{7.04} & \heatdisplaylow{0.0800}{0.0800}{0.4500}{5.51} & \heatdisplaylow{64.3000}{64.2000}{67.4000}{35.7} \\
\bottomrule
\end{tabular}%
}
\end{table*}

\section{Training and Inference Details}
\label{app:labels}
The teacher edit sequence is built online in the dataloader from the canonical draft and reference. Matching is exact, without a voxel tolerance, so the unfiltered set difference reproduces the reference face set exactly. The filter that removes quad diagonal flips (Section~\ref{sec:method:labels}) matches quadruples greedily, does not check convexity, and skips edges shared by more than eight deleted faces. It is bypassed when the set difference deletes more than $2{,}048$ faces, so that drafts with many non-manifold edges cannot stall a dataloader worker, and such pairs are supervised on the exact set difference. The filtered teacher sequence reproduces the reference exactly on $14.3\%$ of held-out pairs, versus all pairs without filtering; thus most pairs differ by a quad diagonal.

The base recipe uses $84\%$ draft/reference pairs, $15\%$ partial-edit pairs, and $1\%$ identity pairs. A partial-edit pair applies a random subset of the teacher edit sequence, and the remaining edits become the label, which matches the input of later correction rounds. The reconstruction pairs are drawn at a continuous noise level (Section~\ref{sec:exp:setup}). Patch hole punching removes further faces in connected patches from $70\%$ of the draft/reference and partial-edit pairs before the teacher edit sequence is built. Every sample is transformed by one of the $48$ rotations and reflections of the voxel grid, applied identically to draft and reference. Patch hole punching starts at base-model step $97{,}000$.

\noindent\textbf{Optimization.} $\mathcal{L}_\mathrm{del}$ averages the deletion-head cross-entropy over all deletion steps in a packed batch, including the end-of-del step, and $\mathcal{L}_\mathrm{add}$ averages the addition-head cross-entropy over all vertex selections, end-of-add decisions, and new-vertex coordinates. Normalizing each phase by its own decision count keeps the addition phase, with at least three decisions per face, from outweighing the deletion phase. The prefix carries no loss. The transformer has $32$ layers of width $2304$ with $18$ heads. Training uses MuonAdamW---Muon for two-dimensional weight matrices and AdamW otherwise---with a constant $4\times10^{-4}$ learning rate after $2$k warmup steps, weight decay $0.1$, gradient clipping $1.0$, bf16, and a fixed per-rank token budget on A800 GPUs.

\noindent\textbf{Decoding safeguards.} Deletions and additions are each capped at $\lfloor F/2\rfloor+64$. An added face that duplicates an existing face is fed back but left out of the edit sequence and does not count toward the cap. A run of $32$ consecutive duplicates or the first degenerate face ends the addition phase. After the edit sequence is applied, the mesh is re-canonicalized. Repeated correction (Table~\ref{tab:ablation}b) allows further rounds and stops on an empty edit sequence, a round cap, or a repeated edit sequence. For export, kept faces start from their draft winding and added faces are oriented by propagation from oriented neighbors, after which a signed-volume check may flip the whole mesh outward. Components without an oriented seed fall back to standard normal repair. Winding does not affect any reported metric.

\end{document}